\documentclass[%
 reprint,
nolongbibliography,
 amsmath,amssymb,
 aps,
10pt]{revtex4-2}

\usepackage[T1]{fontenc}
\usepackage{mathptmx}
\DeclareMathAlphabet{\mathcal}{OMS}{cmsy}{m}{n} % but do not change the mathcal version 
\usepackage{multirow}
\usepackage[font=small]{caption}
\usepackage{float}
\usepackage{mathrsfs}
\usepackage{color}
\usepackage{graphicx}% Include figure files
\usepackage{subeqnarray}
\usepackage{dcolumn}% Align table columns on decimal point
\usepackage{bm}% bold math
\usepackage[%showframe,%Uncomment any one of the following lines to test 
scale=0.82, marginratio={1:1, 1:1}, ignoreall,% default settings
]{geometry}

\begin{document}

\preprint{APS/123-QED}

\title{\large{Routes to Stratified Turbulence Revealed by Unsupervised Classification of Experimental Data}}

\author{Adrien Lefauve}
\email{lefauve@damtp.cam.ac.uk}

\author{Miles M. P. Couchman}

\affiliation{Department of Applied Mathematics and Theoretical Physics, University of Cambridge \\ Centre for Mathematical Sciences, Wilberforce Road, Cambridge CB3 0WA, UK %
\vspace{0.5cm}}%

\date{4 May 2023}% It is always \today, today,
             %  but any date may be explicitly specified

\begin{abstract} 
Modeling fluid turbulence using a `skeleton' of coherent structures has traditionally progressed by focusing on a few canonical experiments, such as pipe flow and Taylor-Couette flow. We here consider an alternative canonical experiment, the stratified inclined duct, a sustained shear flow whose density stratification allows for the exploration of a wealth of new coherent and intermittent states at significantly higher Reynolds numbers than in unstratified flows. We  automatically identify the underlying turbulent skeleton of this experiment with a novel data-driven method combining dimensionality reduction and unsupervised clustering of shadowgraph visualizations. We demonstrate the existence of multiple types of energetic turbulence across parameter space, as well as intermittent turbulence that cycles between these types, revealing distinct transition pathways. Our method and results pave the way for new reduced-order models of multi-physics turbulence.
\end{abstract}

\maketitle

\textit{Introduction.} --- One of the greatest challenges in fluid dynamics is to identify a simple, low-dimensional description of the high-dimensional and strongly nonlinear dynamical system that characterizes turbulence. One promising approach, coherent-structure modeling, dates back to \cite{hopf_mathematical_1948,ruelle_nature_1971} and assumes that the turbulent dynamics of practical importance are low-dimensional, with phase-space trajectories spending significant periods of time near a relatively small set, or `skeleton', of exact (though usually unstable) solutions of the Navier-Stokes equations called simple invariant solutions or exact coherent states \cite{kawahara_significance_2012,graham_exact_2021}. This description would enable the prediction of turbulent statistics using a weighted average over these solutions \cite{cvitanovic_recurrent_2013}, and could also reproduce some aspects of the spatiotemporal coherence of the fully-nonlinear dynamics \cite{lucas_irreversible_2017,page_revealing_2021}. 

The search for such a coherent turbulent skeleton requires a deep understanding of the mechanisms of transition from laminar to turbulent flow. The study of hydrodynamic stability dates back to Osborne Reynolds \cite{reynolds_experimental_1883}, who introduced in his 1883 seminal paper a canonical experiment that has since become a dominant paradigm: cylindrical pipe flow. A key insight was that the transition to turbulence was governed by a dimensionless group of parameters, now known as the Reynolds number $Re=uh/\nu$, where $u$ and $h$ denote the flow's characteristic velocity and length scales and $\nu$ denotes the fluid's kinematic viscosity. Recent reviews \cite{eckhardt_turbulence_2007,willis_experimental_2008,barkley_theoretical_2016,avila_transition_2023} highlight the significant progress made on understanding the route to turbulence and the skeleton of pipe flow in the last 150 years. 
A key step in this journey was made by G. I. Taylor \cite{taylor_stability_1923}, who introduced in 1923 a second canonical experiment: the flow between two concentric rotating cylinders, now known as the Taylor-Couette flow.  
Over the last 100 years, the contrast between the pipe and Taylor-Couette flows revealed two fundamentally different routes to turbulence with increasing $Re$. Pipe flow is linearly stable, and nonlinearities amplify finite-amplitude disturbances (a subcritical transition) into turbulent puffs and slugs of increasing lifetime like in excitable and bistable media \cite{barkley_theoretical_2016}. By contrast, Taylor-Couette flow (when dominated by inner-cylinder rotation) is linearly unstable, and nonlinearities lead to the saturation of exponentially-growing instabilities (a supercritical transition) meaning that turbulent chaos is reached after a sequence of  successive instabilities \cite{coles_transition_1965,feldman_routes_2023}. 

\begin{figure}[t]
\includegraphics[width=0.99\linewidth]{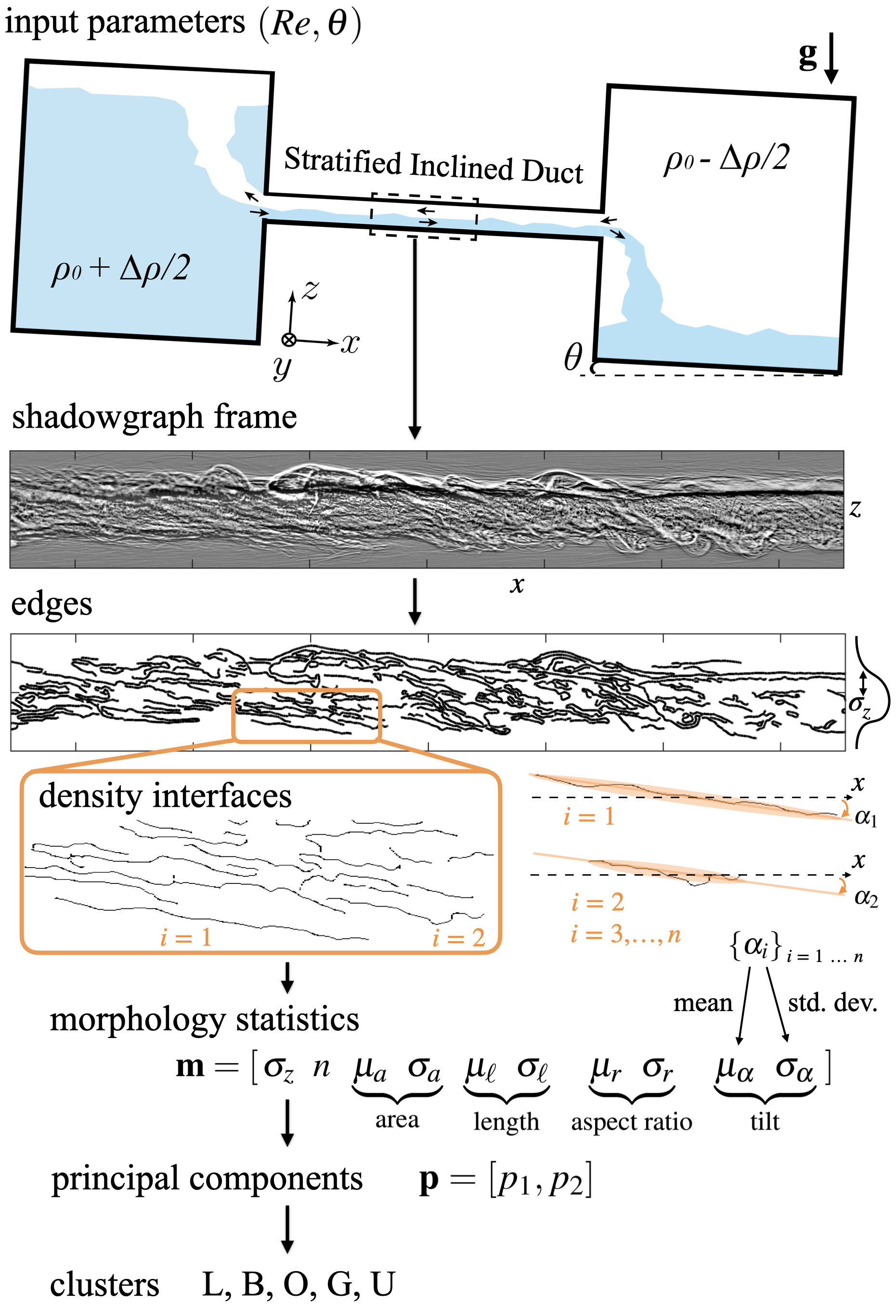}
\caption{Experimental setup (the duct is 2000 mm long, 100 mm wide and $H=50$ mm tall), shadowgraph data and automated dimensionality reduction pipeline: edge detection, 10-dimensional morphology vector, two-dimensional vector in principal coordinates, and unsupervised clustering.}
\label{fig:pipeline}
\end{figure}

We here consider routes to turbulence in a third and comparatively less-well-known laboratory experiment: the `stratified inclined duct' (SID, see top of figure~\ref{fig:pipeline}). This  density-stratified flow features rich transitional and intermittent dynamics which, we will argue, open up a new fruitful paradigm for advancing turbulence modeling. Previous progress in characterising a range of canonical flows has crucially relied upon the study of transitional flows exhibiting spatio-temporal turbulent intermittency. Density stratification introduces stabilizing effects and thus a second dimensionless parameter, allowing this intermittency to be more generic and explored at higher $Re$ \cite{turner_buoyancy_1973,deusebio_intermittency_2015},  %. The additional modes of motions due to the restoring buoyancy force give rise to much richer dynamics, 
thus giving rise to more interesting building blocks for the skeleton of turbulence \cite{lucas_layer_2017,salehipour_deep_2019,smith_turbulence_2021}. 

The first conceptual leap of adding a two-layer stratification to the shear flow in a tilted pipe was already made by Reynolds in the 1883 paper \cite[\S\,12]{reynolds_experimental_1883}, and allowed him to note that the turbulent transition was fundamentally different to pipe flow. The second -- and much later -- conceptual leap leading to the SID experiment \cite{macagno_interfacial_1961,meyer_stratified_2014} was to connect a long rectangular duct to two large reservoirs of saltwater of different densities  $\rho_0 \pm \Delta\rho/2$ (figure~\ref{fig:pipeline}). SID flow has two variable control parameters: $Re$ based on half the duct height $h=H/2$ and the layer-averaged buoyancy velocity scale $u=\sqrt{gH\Delta\rho/\rho_0}$, and the tilt angle $\theta$, which provides extra energy \cite{lefauve_regime_2019} to sustain dissipative states for long periods of time and explore their spatio-temporal intermittency. This `intermittent regime' was described and mapped by \cite{meyer_stratified_2014,lefauve_regime_2019,lefauve_buoyancy_2020} in intermediate regions of parameter space $(Re,  \theta)$, between the finite-amplitude Holmboe wave regime at low values of $Re \, \theta$ (which is mostly laminar with little interfacial mixing) and the fully turbulent regime at high $Re \, \theta$ (which has intense mixing and never relaminarises). Novel experiments that measured the time-resolved, three-dimensional velocity and density field \cite{partridge_versatile_2019} allowed for the preliminary development of the `SID skeleton', demonstrating that (i) the transition to turbulence was supercritical, mediated by the `confined' Holmboe instability \cite{lefauve_structure_2018}, and (ii) some of the three-dimensional coherent structures in the fully turbulent regime  (e.g. hairpin vortices) could be traced back to this linear instability \cite{jiang_evolution_2022}.

In this Letter and companion Article \cite{lefauve_data-driven_2023}, we present a novel data-driven framework that significantly advances the discovery and understanding of the building blocks underpinning SID turbulence across the $(Re,\theta)$ space. This analysis had traditionally relied upon a trained human eye identifying qualitative flow features across entire experimental videos \cite{meyer_stratified_2014,lefauve_buoyancy_2020}, limiting the accuracy, repeatability, and feasibility of classifying large datasets, and hence its success to date. We now overcome these limitations with a method that combines dimensionality reduction and unsupervised clustering to  classify instantaneous flow snapshots objectively. We identify a variety of distinct types of turbulence by physically interpreting the detected clusters, and we reveal unique routes to these turbulent states by studying the distribution of clusters in $(Re,\theta)$ as well as the temporal cycles between them in the intermittent regime.

\textit{Dataset and dimensionality reduction.} --- Our dataset consists of 50,155 individual shadowgraph frames (see example in figure~\ref{fig:pipeline}) belonging to 113 movies. Each movie visualizes the evolution of a sustained, sheared, salt-stratified turbulent flow in SID  for a fixed $(Re,\theta)$ over hundreds of advective time units $h/u$.
%of $O(10)$ duct transit times 
Collectively, the movies span the Holmboe wave, intermittent and fully turbulent human-classified regimes \cite{lefauve_buoyancy_2020} across a wide region of the parameter space $Re=300-5000$, $\theta=1-6^\circ$. %, which will remain out of reach of numerical simulations for the foreseeable future \cite{zhu_stratified_2023}. 
The grayscale intensity in shadowgraphs is approximately proportional to the $x-z$ curvature of the fluid's refractive index, and hence of the density field, integrated over the spanwise direction $y$ (the path of the light rays). The dataset is thus particularly suited to studying the structure and temporal evolution of density interfaces embedded within the flow, structures that are known to be energetically and dynamically meaningful \cite{linden_mixing_1979, caulfield_layering_2021,couchman_mixing_2023}.

Our automated dimensionality reduction pipeline (sketched in figure~\ref{fig:pipeline}) starts with detecting edges in the shadowgraph images using a Canny algorithm \cite{canny_computational_1986}. Second, individual density interfaces are extracted as sets of connected `edge' pixels, and their properties are computed: number per frame $n$, lists of respective areas $\{a_i\}_{i=1,\ldots,n}$,  lengths $\{\ell_i\}$, aspect ratios $\{r_i\}$ and tilt angles $\{\alpha_i\}$ based on the fitting of an ellipse (sketched in orange). Third, the 10-dimensional morphology vector
\begin{equation}\label{eq:m}
    \mathbf{m}_f=[ \sigma_z \ \  n \ \ \mu_a \ \ \sigma_a \ \ \mu_\ell \ \ \sigma_\ell \ \ \mu_r \ \ \sigma_r \ \ \mu_\alpha \ \ \sigma_\alpha]
\end{equation}
is constructed to represent each frame $f$ by its number of interfaces $n$, the vertical standard deviation of edge pixel density $\sigma_z$, and the mean ($\mu$) and standard deviation ($\sigma$) of its four lists of interface properties. Fourth, a principal component analysis (PCA) \cite{brunton_data-driven_2019} is performed to take advantage of correlations in the data matrix $\mathbf{M}=[\mathbf{m}_f]_{f=1,\ldots,50155}$, and the PCA basis is truncated at second order, explaining 80\,\% of the variance. This yields a rank-two data matrix $\mathbf{P}=[\mathbf{p}_f]_{f=1,\ldots,50155}$ where each  1.5-MPixel shadowgraph frame has been compressed into a two-dimensional vector $\mathbf{p}_f$ in PCA space $(P_1,P_2)$.

\begin{figure}[t]
\includegraphics[width=\linewidth]{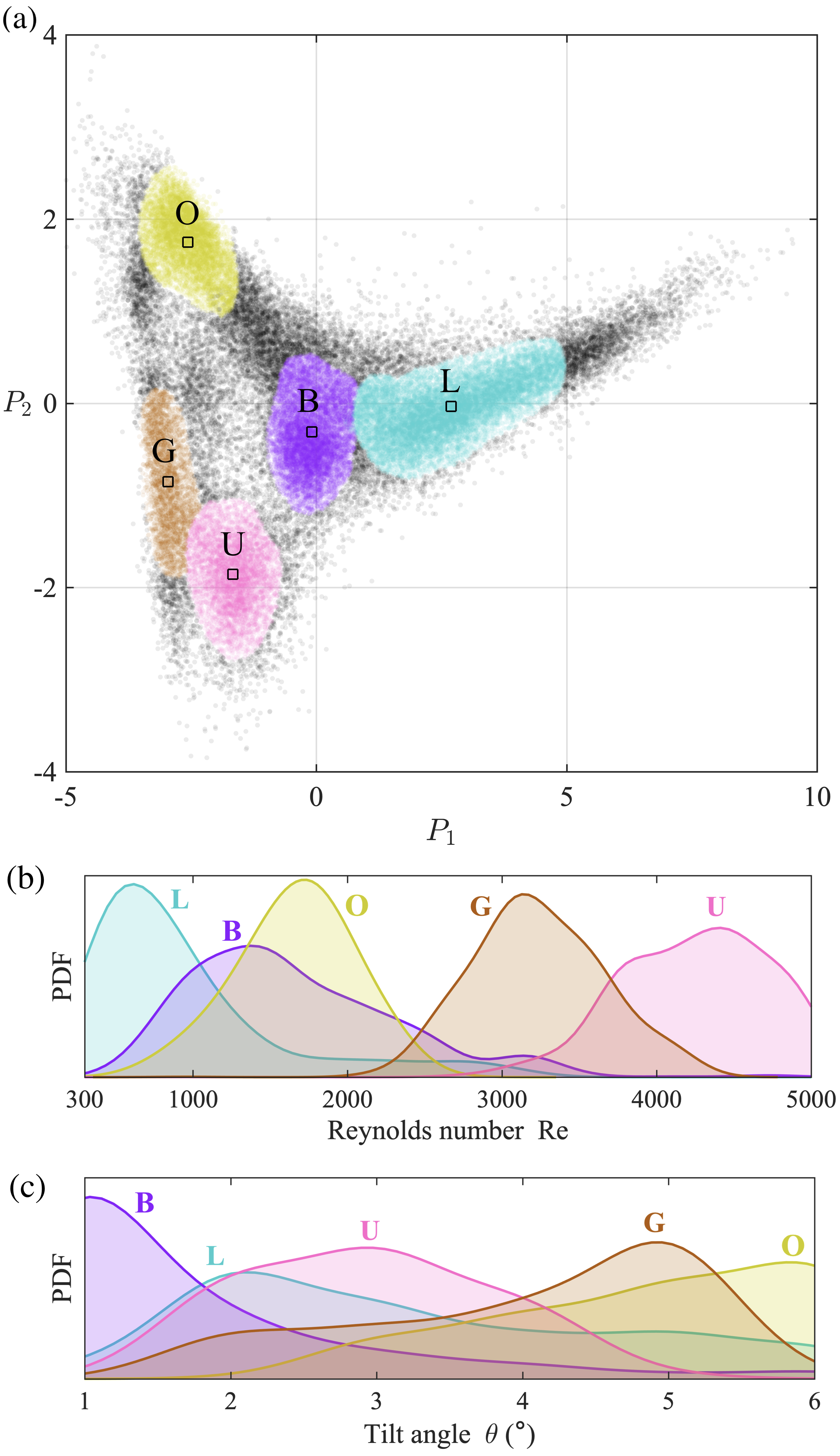}
\caption{Clustered data:  All 50155 experimental frames in the space of principal components (a). Points belonging to clusters are colored, while sparser, unclustered points are in black. Smoothed probability distributions (PDFs) of clusters in $Re$ (b) and $\theta$ (c). }
\label{fig:clusters}
\end{figure}

\textit{Classification and physical interpretation.} -- The  density-based clustering algorithm `OPTICS' \citep{ankerst_ordering_1999,couchman_data-driven_2021} is then applied to $\mathbf{P}$ to automatically determine a natural grouping of dense regions in PCA space. Figure~\ref{fig:clusters}a shows that five clusters are detected (number automatically determined), accounting for nearly 80~\% of all data points, with unclustered points lying in sparser regions. Figures~\ref{fig:clusters}b-c show the distributions of $Re$ and $\theta$ for the frames belonging to each cluster. The successive prevalence of each cluster L, B, O, G and U as $Re$ increases, and of a different succession of clusters as $\theta$ increases, already suggests that the classification captures meaningful and non-trivial physics. A complete description of the ratio of time spent in each cluster in all 113 experiments spanning the $(Re,\theta)$ space is given in the companion Article \cite{lefauve_data-driven_2023}.

\begin{figure}[t]
\includegraphics[width=0.97\linewidth]{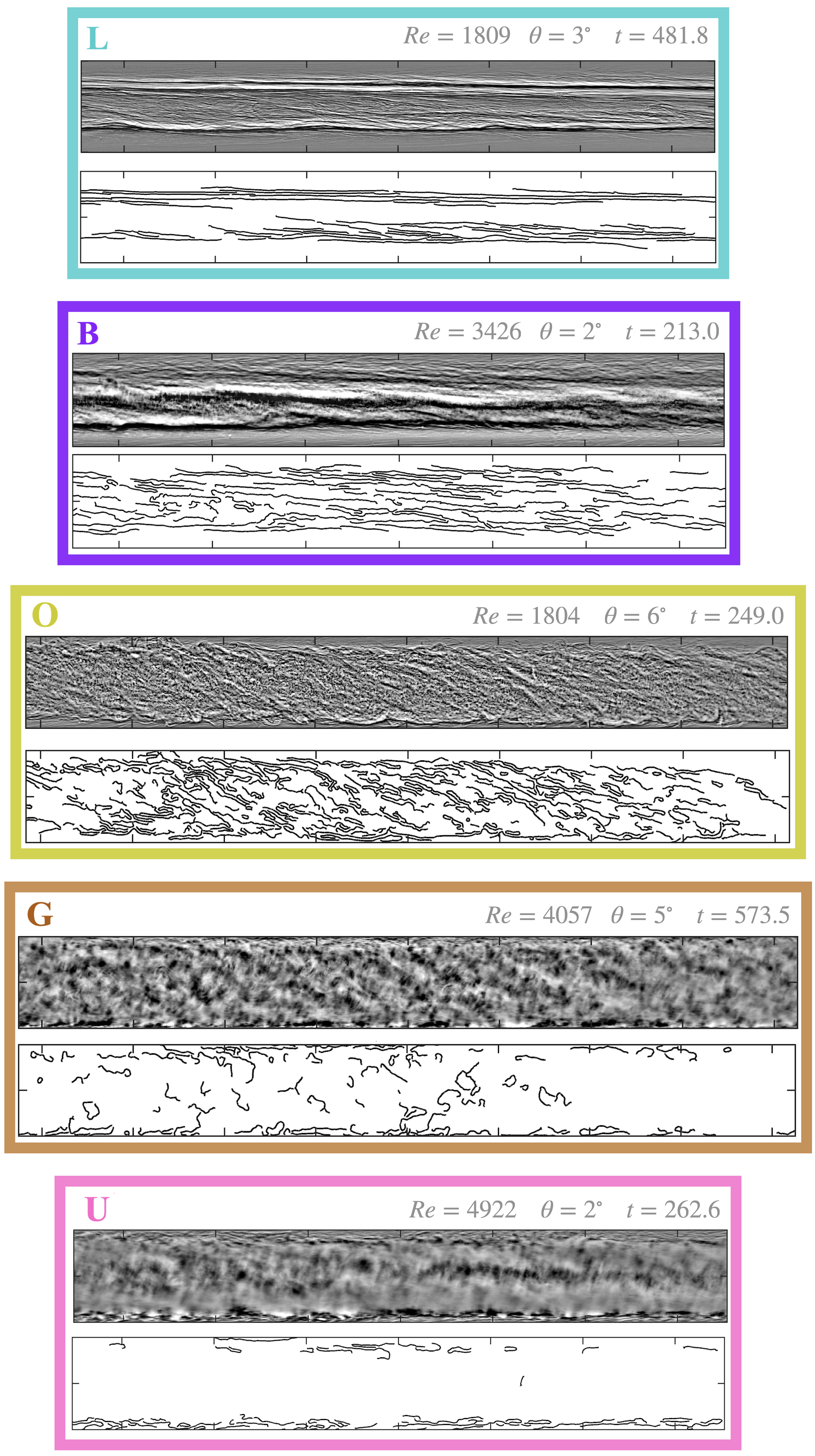}
\caption{Frames corresponding to the centroid of each cluster (squares in figure~\ref{fig:clusters}a) representing the five types of turbulence. Shadowgraphs are shown (top) with the matching edges (bottom). Note the dimensionless $Re,\theta$ and snapshot time $t$ (top right).}
\label{fig:examples}
\end{figure}

We now interpret these clusters in figure~\ref{fig:examples} by considering representative frames corresponding to the centroid of each cluster. We show the original shadowgraphs (top) with the detected density edges (bottom), effectively `inverting' our dimensionality reduction pipeline. Frame L intuitively qualifies as the most laminar state (for reference, a hypothetical single laminar interface would yield principal coordinates far off the top right vertex in figure~\ref{fig:clusters}a). It represents what we may call \textit{laminarizing turbulence}, where small-scale structure from a preceding turbulent phase remains visible in the shadowgraph snapshot, but the pattern of edges shows a `stacking' of flat and stable density interfaces. By contrast, frame B illustrates \textit{braided turbulence}, owing to its pair of central `braids' with relatively two-dimensional curvature in the density field evidenced by a strong contrast in shadowgraph intensity, representing two strong interfaces surrounded by a number of weaker ones. Frame O illustrates \textit{overturning turbulence}, owing to its numerous short, tilted and hence unstable interfaces. Frames G and U illustrate \textit{granular} and \textit{unstructured turbulence} respectively, owing to strongly three-dimensional curvature in the density field, blurring the shadowgraph image across most of the duct height. The similarity of clusters G and U is consistent with them being adjacent in figure~\ref{fig:clusters}a; the detailed OPTICS results in \cite{lefauve_data-driven_2023} show that they can in fact be considered sub-components of a larger, unifying cluster. Nevertheless, these two clusters are distinguished by subtle differences in their density interfaces. The turbulence in frame G is more granular, owing to slightly stronger contrast and hence more detected edges, especially at mid-height, whereas it has less structure in frame U, with flatter edges located nearer the top and bottom walls, where the blurred mixing layer meets the more quiescent boundary layers. Further quantitative descriptions of the types of turbulence identified by these clusters are provided in the companion Article \cite{lefauve_data-driven_2023} through the corresponding 10-dimensional vectors of morphology statistics \eqref{eq:m}. %For example, density interfaces in laminarizing turbulence (type L) are typically concentrated in a single set at mid height (low $\sigma_z$), are long (high $\mu_\ell$), slender (low $\mu_r$), and flat (low $\mu_\alpha$). By contrast, density interfaces in overturning turbulence (type O) are spread out in $z$, far more numerous and tilted, but shorter and thicker. We also highlight the robustness of our dimensionality reduction and clustering approach to variations in pixel size between the frames in the dataset (as seen in figure~\ref{fig:examples}). This is especially useful when accumulating data from experimental campaigns with different cameras or fields of view, as done here. 

\begin{figure}[t!]
\includegraphics[width=\linewidth]{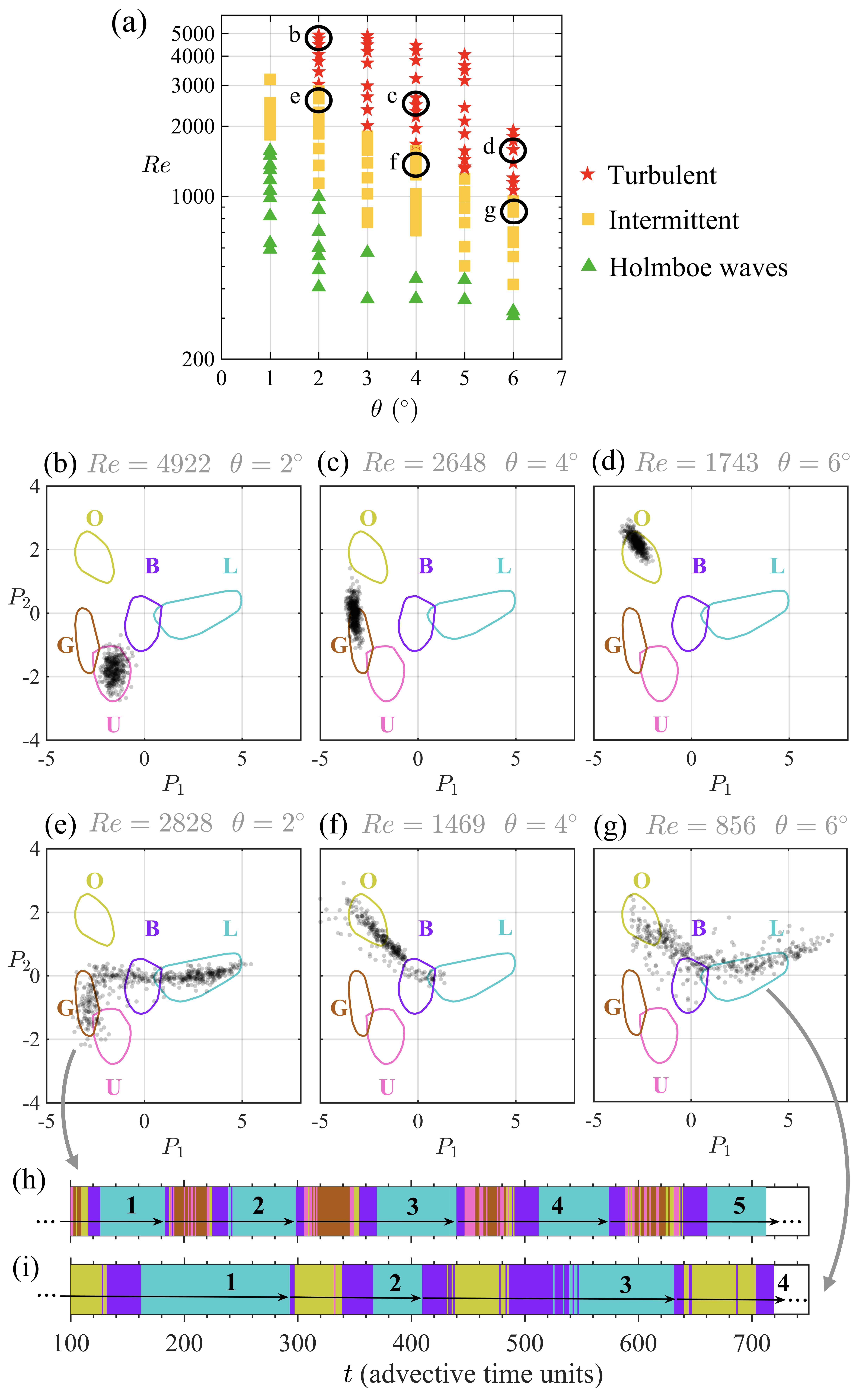}
\caption{Temporal dynamics (b-i) in six experiments chosen from the dataset of 113 experiments in the space of input parameters (a). The frames are shown in (b-g) by transluscent symbols and the cluster boundaries are shown in colors. The cycling time series of (e,g) are shown in (h,i) respectively (initial transients $t<100$ discarded), with cycle numbers in bold. Unclustered frames are colored based on the nearest cluster.}
\label{fig:timeseries}
\end{figure}

\textit{Phase-space dynamics and temporal intermittency.} --- We now examine in figure~\ref{fig:timeseries} the trajectory in phase space $(P_1,P_2)$ of six different experiments, chosen across  $(Re,\theta)$ from the dataset of 113 experiments  (figure~\ref{fig:timeseries}a). We compare three experiments traditionally classified (by the human eye) as fully `Turbulent' at $Re \, \theta \approx 1\times 10^4$ (figures~\ref{fig:timeseries}b-d) and three classified as `Intermittent' at $Re \, \theta \approx 6\times10^3$ (figures~\ref{fig:timeseries}e-g). We note that the product $Re \, \theta$ is proportional to  the time- and volume-averaged rate of turbulent kinetic energy dissipation in the flow $\epsilon$ \cite{lefauve_regime_2019}, and to the dynamic range of stratified turbulence, or buoyancy Reynolds number $Re_b=(L_O/L_K)^{4/3}$, measuring the separation between the Ozmidov $L_O=(\epsilon/N^3)^{1/2}$ and Kolmogorov $L_K=(\nu^3/\epsilon)^{1/4}$ turbulent lengthscales ($N$ is the averaged buoyancy frequency) \cite[Sec. 5.1]{lefauve_experimental2_2022}. The most dissipative turbulence along the top $Re\,\theta$ line of highest dynamic range assumes different types. At high $Re$ and low $\theta=2^\circ$ (figure~\ref{fig:timeseries}b) turbulence is exclusively unstructured, staying within or very near cluster U. At intermediate $Re$ and $\theta=4^\circ$ (figure~\ref{fig:timeseries}c), it shifts to being granular in cluster G with brief excursions to sparser (unclustered) space towards O. At low $Re$ and high $\theta=6^\circ$ (figure~\ref{fig:timeseries}d), it is exclusively of overturning type, in or near cluster O.

% FULL PAGE OPTION 
% \begin{widetext}
%     \begin{minipage}{\linewidth}
%         \begin{figure}[H]
%             \centering
% \includegraphics[width=0.8\linewidth]{figures/timeseries_clusters.png}
%             \label{fig:sample}
%             \caption{ddd}
%         \end{figure}    
%     \end{minipage}
% \end{widetext}

By contrast, less dissipative turbulence along the bottom $Re\,\theta$ line transitions between clusters over time (figures~\ref{fig:timeseries}e-i). At intermediate $Re$ and $\theta$ (figure~\ref{fig:timeseries}f), the turbulence is of mixed braided/overturning type, confined to clusters B, O and the intervening space, with rare relaminarizations. At high $Re$ and low $\theta$ (figures~\ref{fig:timeseries}e,h), the route to turbulence is fundamentally different: the time series is now quasi-periodic across L, B and G. The cycles, numbered in bold, have period $T \approx 120-140$ advective units. Each cycle starts with a very short excursion  to braided turbulence (residence time $\approx 5$), followed by a long visit in or near granular turbulence ($\approx 50$), then another excursion  to braided turbulence ($\approx 10$), eventually leading to a long relaminarization ($\approx 65$). At low $Re$ and high $\theta$ (figures~\ref{fig:timeseries}g,i), the route to turbulence is different again, being quasi-periodic across L, B and O (with a more variable period $T\approx 120-220$). Importantly, the intense turbulent periods are of overturning (O) rather than granular (G) type. However they are again accessed by the same braided (B) `gateway', albeit by `turning up' via the top end of the cluster rather by `turning down' via the middle (compare figures~\ref{fig:timeseries}g and e). Moreover, in both cases the excursions through B during the relaminarization phases G/O$\rightarrow$B$\rightarrow$L are consistently longer than during the unstable transitional phases L$\rightarrow$B$\rightarrow$G/O. These results suggest that the dynamical system of SID intermittency organizes around two inherently different `slow manifolds', motivating the extension to a higher-dimensional phase space to resolve the bursting and relaxation dynamics in the orthogonal `fast manifold' \cite[\S~6.7.2]{schmid_data-driven_2021}. The phases of these cycles are reminiscent of the life cycle of a (transient) 
Kelvin-Helmholtz billow \cite{mashayek_goldilocks_2021,smith_turbulence_2021}, but SID flow harbours a greater wealth of turbulent attractors and transition pathways. 

\textit{Conclusions.} --- We have performed the first coherent structure modeling of turbulence in the stratified inclined duct (SID) experiment, a sustained two-layer shear flow whose density stratification and two-dimensional parameter space $(Re,\theta)$ yield a rich set of turbulent and intermittent states. We developed a novel image-processing algorithm to transform experimental shadowgraph movies into a reduced set of two-dimensional vectors representing the morphology of density interfaces within each frame. This allowed an unsupervised algorithm to automatically reveal and map the distribution in parameter space of five distinct types of turbulence, which we interpreted as building blocks for the coherent `skeleton' underpinning SID turbulence.
%We first reduced the dimensionality of a large experimental dataset of shadowgraph movies by projecting each megapixel snapsot to a physically-interpretable two-dimensional phase space. We were then able to detect clusters in this space with an unsupervised algorithm, allowing us to classify turbulence based on the quantitative morphology of density interfaces. 

The temporal dynamics of turbulence with high dissipation and dynamic range $\propto Re \, \theta$ are confined within individual clusters, whose type shifts from unstructured to granular to overturning with decreasing $Re$ and increasing $\theta$. Less dissipative, intermittent turbulence at lower $Re \, \theta$ transitions between clusters, typically quasi-periodically. Two fundamentally different routes to turbulence were identified, both of which pass through a common braided turbulence `gateway' but end up in different granular or overturning `attractors'.
%with turbulent and relaminarization phases always being accessed through braided turbulence.%, with varying residence times in each cluster. 
%Analysis of these cycles revealed that two fundamentally different routes are connected to this common `braided turbulence gateway': granular and overturning turbulence.
%: at higher $Re$ and lower $\theta= 2^\circ$ the phase space trajectories always curve `downwards' towards granular turbulence, while at lower $Re$ and higher $\theta= 6^\circ$ they always curve `upwards' towards overturning turbulence. 
The presence of at least two transitions within SID highlights the untapped potential of this canonical flow for the broader understanding of high-$Re$ turbulence with extra physics, e.g. rotating, multiphase, or magnetohydrodynamic turbulence. %Our data-driven method can be generalized and assist the discovery of such complex dynamical systems from insightful experiments and simulations.

\acknowledgments{We thank Xianyang Jiang, Gaopan Kong and the technicians of the G. K. Batchelor Laboratory for their help with the experiments. We are also grateful to Paul Linden and  Stuart Dalziel for their support  and to Colm-cille Caulfield for insightful discussions on the implications of this work. The experimental facility was funded by the ERC grant  `Stratified Turbulence And Mixing Processes' (STAMP, No 742480). A.L. acknowledges funding from a Leverhulme Early Career Fellowship and a NERC Independent Research Fellowship (NE/W008971/1). All data will be made available. }

%\bibliography{AL_references_2022_01.bib}% Produces the bibliography via BibTeX.

%apsrev4-2.bst 2019-01-14 (MD) hand-edited version of apsrev4-1.bst
%Control: key (0)
%Control: author (8) initials jnrlst
%Control: editor formatted (1) identically to author
%Control: production of article title (-1) disabled
%Control: page (0) single
%Control: year (1) truncated
%Control: production of eprint (0) enabled
%

\end{document}